# SUPERFLUID HELIUM AS THE CONDENSATE OF BOUND ATOMS PAIRS

**N.M.Blagoveshchenskii**
*Institute of Physics and Power Engineering, 249033, Obninsk, Russia*
E-mail: blagoveshchenskii@rambler.ru

**ABSTRACT.** Based upon the hypothesis of simultaneous participation of helium particles in atom - atom and pair – pair motions, the proof of Feynman and Bogoliubov formulas validity for elementary excitations spectrum in superfluid helium is carried out, with the double mass of particle, 2M. The value of effective mass for such a motion, 1.5M, leads to the more close value 2.09K for $\lambda$ – transition temperature (2.17K), with respect to the value of 3.14K obtained with use of single mass M.

**INTRODUCTION.** The pioneering theory of superfluidity [1] based upon the hypothesis that the majority of liquid helium atoms obeys the condition for Bose – Einstein Condensation (BEC) into the zero momentum state. The experimental investigation of single atom condensate density by means of inelastic neutrons scattering [2] has revealed, however, that the BEC state belongs not to the majority of atoms, but to the part of about 10% only. The further theoretical investigations [3,4] have allowed the possibility of the superfluidity phenomenon to occur just for BEC density of the order ~ 10%. The main disadvantage of these theories was too small value of sound velocity, namely $c \sim \sqrt{n(BEC)}$ with respect to needed $c \sim \sqrt{n}$.

The next step come with the hypothesis of the the possible existence of macroscopic fraction of pair condensate in liquid helium [5,6]. In these theories either the single atom BEC fraction was ignored at all, or the parts of single and pair (and of higher orders) condensates were studied independently [7].

In present work we assume the helium atoms participating in both the single atom – atom and pair – pair motions, thus possessing the independent relaxation times for ground state of liquid helium. In this spirit the pair condensation [5] is not more stationary but commonly dynamical, and the "momental" density of dynamical bose – condensate, in addition with density of conventional single BEC is equal 100%, at least at $T \to 0$.

**1.** Microdynamics of liquid helium from a point of establishing the structure of the ground state can be developed in terms of so-called memory function $\Gamma(t)$, which is the kernel of integro – differential Volterra equation [8] for Velocity Autocorrelation Function (VACF), $\Psi(t)$:

$$\partial\Psi/\partial t = - \int \Gamma(\tau)\Psi(t-\tau)d\tau \quad . \tag{1}$$

Memory function is usually two – component and can be written:

$$\Gamma(Q,t) = \Gamma(Q,0)[\alpha(Q)\exp(-t/\tau(\mu)) + (1- \alpha(Q))\exp(-t/\tau(\alpha))] \quad , \qquad (2)$$

here $\tau(\mu)$ and $\tau(\alpha)$ – times of fast (single atom – atom) and slow (corresponded to the nearest surrounding decay) relaxation of atom or cluster with momentum Q, and $\alpha(Q)$ and $(1- \alpha(Q))$ – weights of slow and fast components of $\Gamma(t)$.

Assuming that the minimal cluster of slow relaxation is tetrahedron constructed from two pairs of interactive atoms, one can estimate [9] the ratio of densities k for atoms taking place both pair – pair and atom – atom motions:

$$k = 2 \tau(\alpha)/ \tau(\mu)[1- \alpha(Q)] \qquad (3)$$

As shown in [9], for liquid potassium k = 0.2; for liquid sodium k = 0.4; for liquid lithium k = 1.0. Thus, lithium can be regarded as liquid, those excitation spectrum is forming as in single – atom, both in two – atom densities. The analogues picture we have in liquid helium. As shown in [9], time of fast relaxation $\tau(\mu)$ is equal to the decay time $\tau$, inferred from the VACF analysis by the decaying cosine formula [10]:

$$\Psi(t) = \mathrm{sech}(t/ \tau)\cos(\omega t) \quad , \qquad (4)$$

By analogy, we can propose that for slow relaxation time $\tau(\alpha)$ one can detect the same resemblance in two – component model (4), and for the case of liquid helium we hold the analysis of VACF $\Psi(t)$ instead of direct analysis of memory function $\Gamma(t)$.
VACF of liquid helium for T = 3K has been obtained with use of standard molecular-dynamical program MOLDY [11] (see fig.1) and demonstrated sharply two – component structure. Times relaxation ratio $\tau(\alpha) / \tau(\mu)$ is nearly 5, while components areas ratio $N(\alpha)/N(\mu) = 0.61$.
Not numerous data [12] of liquid helium VACF analysis at other temperatures shows k ~ 0.1 at T = 4K, and, likely k = 1.0 near $\lambda$ – point. Thus, atoms of normal liquid helium I take place both in atom – atom, and pair – pair motions with densities ratio $N(\alpha)/N(\mu)$, varying from 0 (at boiling point T = 4.2K) to about 1 (at $\lambda$ – point T = 2.17K).

Propose now, that at the interval from 0 up to 2.17K, the ratio $N(\alpha)/N(\mu) = \mathrm{const} = 1$, but the ratio of bound pairs number to the number of total interacting pairs is varying. So the $\lambda$ – point can be determined as the temperature where the pairs with non-zero bound constant in helium II are appearing. Perhaps they produce the superfluid component, and their energy at T = 0 is equal to the zero – oscillations energy $\Delta$.

**2.** Namely this pattern is reproduced by paper [13], where the possible existence of bound pairs appears due to off-diagonal part of interaction Hamiltonian:

$$U(\Lambda) = 1/2V\sum v(k)[a^{\times}(k)a^{\times}(-k)a(0)^2+a^*(0)^2 a(k)a(-k)] \qquad (5)$$

Bound pairs are the cause of appearance of so-called dynamical bose-condensation for pairs with zero-total momentum, and the bounding constant is defined as follows:

$$g(\Lambda,00) = -1/4(2\pi)^3 \int d^3k \, [v(k)]^2/\varepsilon(k) , \qquad (6)$$

here $\varepsilon(k) = k^2/2M$, and $v(k) = \int d^3x \, \varphi(x) \exp(-ikx)$ – Fourier-transform of potential.

Then we take $v(k)$ from the Bogoluibov theory [1] calculated with experimental form of elementary excitation spectrum $\varepsilon(k)$ [14]:

$$\varepsilon(k)^2 = [k^2/2M]^2 + k^2 nv(k)/M . \qquad (7)$$

and get $g(\Lambda,00) = 0.135$ meV.
On the other hand, executing the numerical estimation by formula (6), we suppose that all the atoms of density n, taking place in creation of the bound pairs, have the same modulus of momentum k (T ~ 0 situation). Calculating the value of $g(\Lambda,00)$, we let the level of zero oscillations near T = 0 equal 1.1 meV[15]. Then we get $g(\Lambda,00) = 0.04$ meV, - the value 3 times less than straightforward estimation of $g(\Lambda,00)$ by formula (6). This contradiction can be understood by the fact that really the renormalized potential $v(k)$ takes place, leading to the modification [5] of initial part in $v(k)$, thus leading to the reduced value of integrand (6). This leads to the correct evaluation [13] of E(b) sign:

$$E(b) = \tfrac{1}{2}v(0) + g(\Lambda,00) , \qquad (8)$$

where $v(0)$ – value of $v(k)$ at k = 0. Bounding rule reads $E(b) < 0$.

We have pointed out that the value of bounding constant is much less than level of zero oscillations $\Delta$, and each atom of interacting pair possesses ~ $\tfrac{1}{2}\Delta = 0.55$ meV energy. Average (effective) temperature of zero oscillations level is $<T> = 0.45$ meV, with respect to existence of single particle BEC participation ( ~ 10% n) in the kinetic energy. This value of $<T>$ will be used below for calculation of one-atom and two-atom densities in S(Q) correction for demonstration of Feinman formula [16] validity,

$$\varepsilon(Q) = Q^2/[2MS(Q)] , \qquad (9)$$

which appears to work at *double* mass of helium atom M → 2M in wide Q area .

   **3.**  Indeed, as is shown in Fig.2, formal setting of structure factor S(Q) into expression (9) reveals for M → 2M the remarkable proximity of calculated $\varepsilon(Q)$ with experimental spectrum, in the area of Q near roton minimum. Now let us demonstrate the marked coincidence in the phonon area as well, with thorough correction of S(Q).
   Notice that at small momenta, S(Q) is equal to the intensity of neutron scattering by helium, Z(Q). Moreover, assuming the experimental Z(Q) [17] to be consisted of scattering part by one-atom density, and part of scattering by two-atom density (the dynamical bose-condensate), let us write these parts separately as:

$$Z(1) = 4\pi Q^2 \exp[-cQ/<T>] \tag{10}$$

and $$Z(2) = 4\pi Q^2 \exp[-\varepsilon(Q)/<T>] \quad , \tag{11}$$

where $cQ$ – phonon part of spectrum $\varepsilon(Q)$. The sound velocity $c$ is the same both in one-atom, and in two-atom densities and can be expressed in small wavenumber limit as

$$c = \sqrt{[1/M(\partial P/\partial \rho)]} \quad , \tag{12}$$

where P is pressure, and $\rho$ – density.

Fig. 3 demonstrates that neutron scattering by one-atom density leads to the reduction of the Z(Q) slope by factor of 2 and exhibits $\partial Z(2)/\partial Q = \partial Z(1)/\partial Q = 1/(4Mc)$, while experimentally observed (summary) slope (curve 3) is $\partial Z/\partial Q = 1/(2Mc)$, and calculated curve (3) nicely accords with experimental [13] Z(Q), revealing the so-called prepeak in Z(Q) and S(Q), at Q ~ 0.5.
   The curve (2) exhibits the correction Z(1) (expression (10) with effective energy or temperature $<T> = 0.45$ meV). This correction needed to be extracted from S(Q) to get final structure factor, shown in Fig. 4 (circles). It is of interest that its slope is $1/(4Mc)$, not $1/(2Mc)$, what demonstrates the S(Q) to be formed indeed by pair-pair interaction, and that the Feynman formula (9) works correctly with *double* mass $M \rightarrow 2M$. The result of calculation is shown at Fig. 5. It demonstrates the excellent agreement with experimental excitation spectrum [14].
   It should be noted also the other attempt [6] to transform the structure factor for demonstration of Feynman formula validity just for $M \rightarrow 2M$. However, paper [6] is based on changing of S(Q) by the value $\Delta S(Q) = const(Q) = S(0)$ but not on the hypothesis of one-atom component Z(1) participation in the observed S(Q). It must be noted that formally the agreement of calculation using theory [6] with neutron experiment on $\varepsilon(Q)$ [14] seems to be good, and shows the appropriate resemblance in both roton and phonon areas of the elementary excitation spectrum. It must be underlined that paper [6], as far as theory [5] operates in terms of *stationary* pair condensation, while our approach and theory [13] work in terms of *dynamical* pair condensation, simultaneously with existence of conventional BEC ( ~ 10%).

   **CONCLUSION.**   Relying upon the hypothesis of existence in superfluid helium the dynamical condensate of bound atoms pairs, one can affirm the validity for Bogoliubov theory [1], demanding the particles state in condensed majority. Such the condensate of bound atoms pairs together with traditional one-atom BEC determines the ground state for superfluid helium system. Also, the validity of Feynman formula (9) [16] is established for $M \rightarrow 2M$ and with thorough correction of S(Q) on the one-atom part of motion in unperturbed density. It is interestingly that contemporary investigation of excitation helium spectra [19] shows in fact the existence of two-component phonon picture, thus affirming the two-component structure of memory function, leading to existence in superfluid helium of the dynamical condensate of bound atoms pairs. Certainly, namely bound atoms pairs together with ordinary one-atom BEC forms the superfluid component of liquid helium II: its density at absolute zero temperature is 100%; at λ – point 0%, accordingly to density of BEC equal zero at T > 2.17K.

Based on fact of two-component memory function below λ – point, it is seen that reduced effective mass of helium atom is 1.5M. So the conventional estimation of λ – transition temperature leads to the 1.5 times less value than one with single mass, namely:

$$T(\lambda) = 3.14/1.5 = 2.09K$$ – number more close with 2.17K than 3.14K.

It is pleasure to thank A.G.Novikov and A.V.Puchkov for instant interest in this work; V.V.Savostin for technical assistance and A.V.Mokshin for valuable remark.

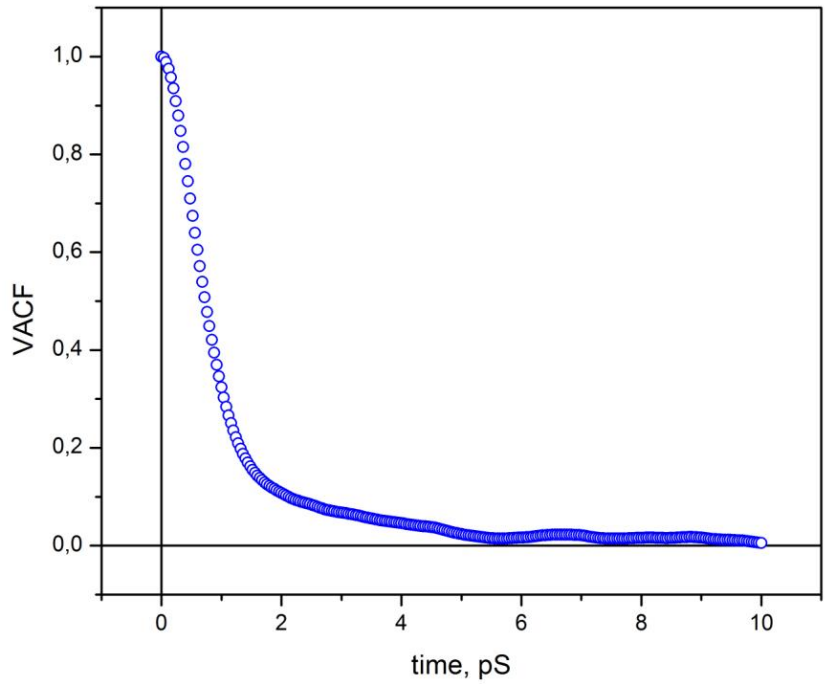

Fig. 1 Liquid helium VACF at T = 3.0K obtained by the use of molecular – dynamical program MOLDY [11]. Calculation by two-component formula (4) shows the areas ratio (slow to fast) k = 0.61.

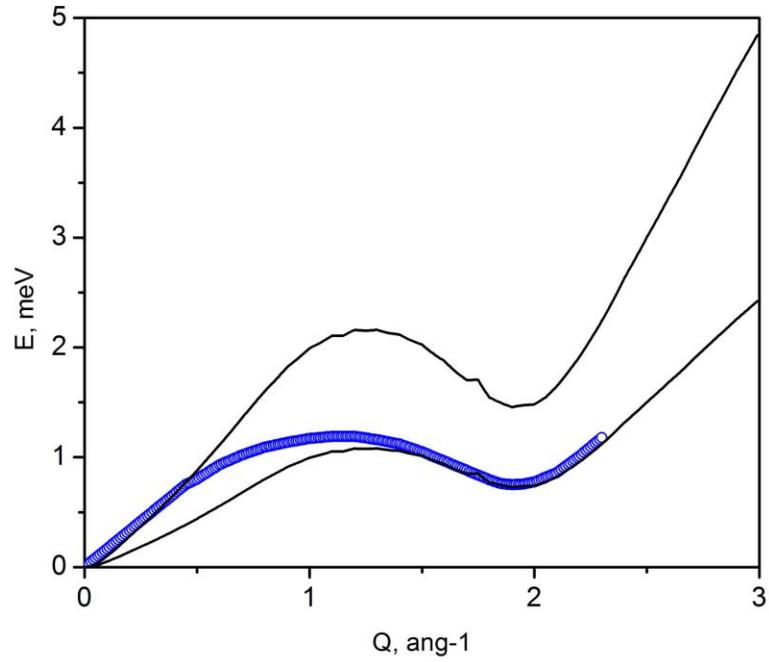

Fig. 2  Excitation spectrum calculated by Feinman formula (9) with single (upper curve) and double mass (lower curve). The structure factor S(Q) is from the paper [18]. Mark the coincidence with experimental $\varepsilon(Q)$ [14] near roton minimum, for $M \to 2M$.

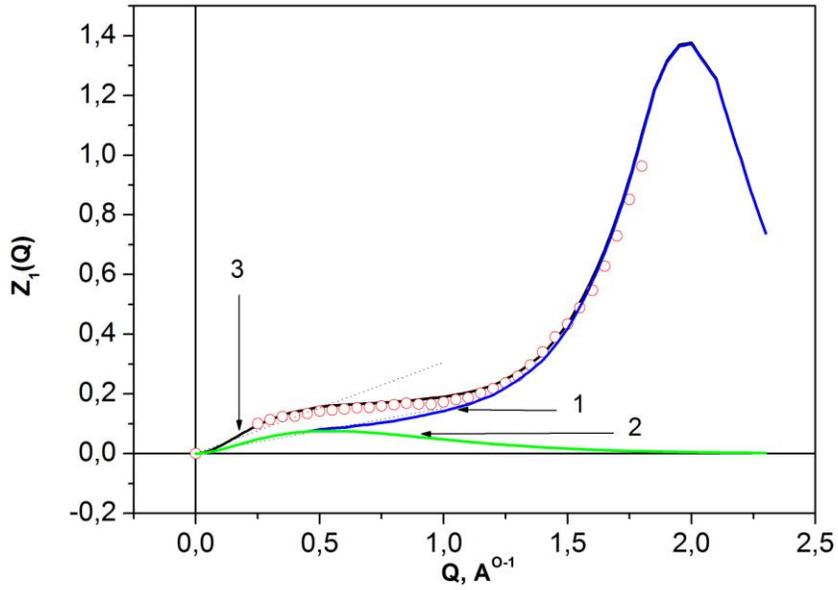

Fig.3 (1) – Scattering by two-atom density Z(2) (formula 11); (2) – scattering by one-atom density Z(1) (formula 10); (3) – summary neutron scattering Z(Q) in comparison with experimental [13] (circles).

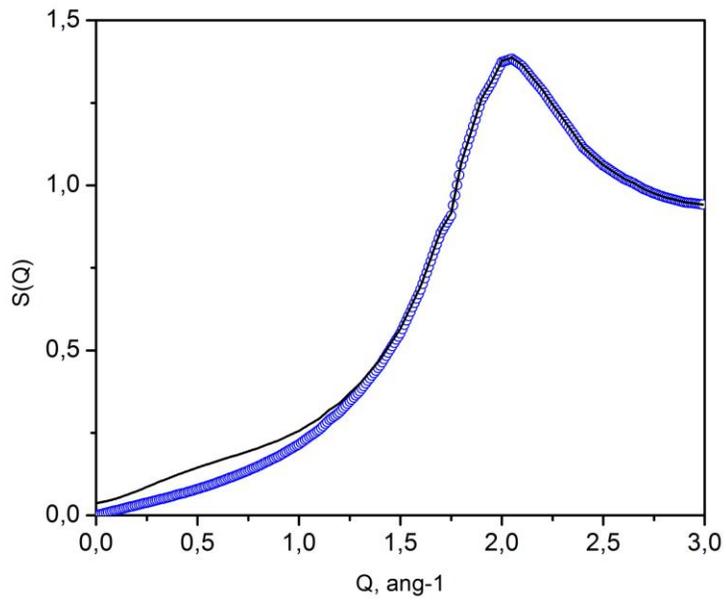

Fig. 4    Structure factor S(Q) [18] – solid line. Circles: S(Q) – Z(1). Slope at Q → 0 is 1/(4Mc). Pay attention that the corrected S(Q) exhibits no any prepeak at Q ~ 0.5.

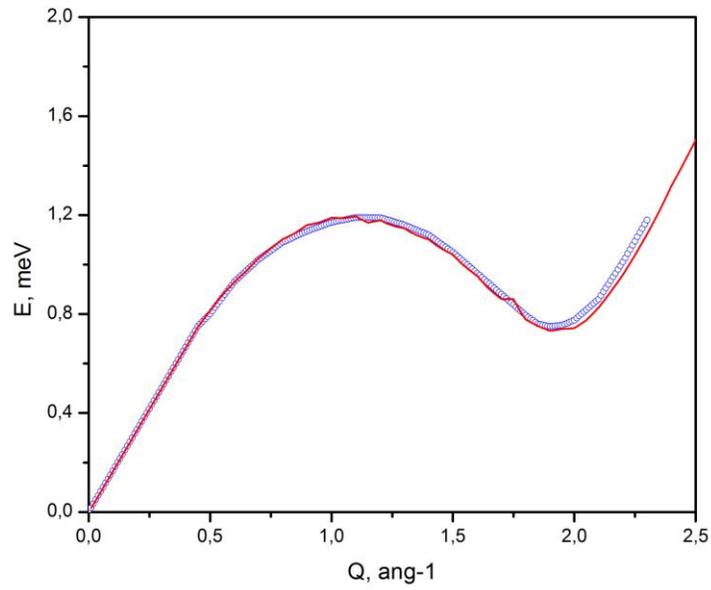

Fig. 5   Elementary excitation spectrum of helium II, calculated by expression (9) with structure factor S(Q) corrected for one-atom part Z(1) by formula (10) and double mass. Circles – the experimentally observed spectrum [14].